\documentclass[12pt,a4paper]{article}
\usepackage{ifthen} % for conditional statements
\newboolean{pdflatex}
\setboolean{pdflatex}{true} % False for eps figures 

\newboolean{articletitles}
\setboolean{articletitles}{true} % False removes titles in references

\newboolean{uprightparticles}
\setboolean{uprightparticles}{false} %True for upright particle symbols

\def\paperauthors{LHCb collaboration} % Leave as is for PAPER, CONF and FIGURE
\def\paperasciititle{Improved measurement of CP violation parameters in Bs2JpsiKK decays} % Set ASCII title here !! MAKE sure it's only ASCII characters !! 
\def\papertitle{Improved measurement of \CP violation parameters in $\Bs\to\Jpsi\Kp\Km$ decays in the vicinity of the $\phi(1020)$ resonance} % Latex formatted title
\def\paperkeywords{{High Energy Physics}, {LHCb}} % Comma separated list
\def\papercopyright{\the\year\ CERN for the benefit of the LHCb collaboration} % new since 9/Apr/2018
\def\paperlicence{CC BY 4.0 licence}
\def\paperlicenceurl{https://creativecommons.org/licenses/by/4.0/}

\usepackage[T1]{fontenc}
\usepackage[top=1in, bottom=1.25in, left=1in, right=1in]{geometry}

\usepackage{microtype}
\usepackage{lineno}  % for line numbering during review
\usepackage{xspace} % To avoid problems with missing or double spaces after
\usepackage{caption} %these three command get the figure and table captions automatically small

\usepackage{graphicx}  % to include figures (can also use other packages)
\usepackage{color}
\usepackage{colortbl}
\usepackage{tabularx}
\graphicspath{{./figs/}} % Make Latex search fig subdir for figures
\DeclareGraphicsExtensions{.pdf,.PDF,png,.PNG}

\usepackage{amsmath,bm} % Adds a large collection of math symbols
\usepackage{amssymb}
\usepackage{amsfonts}
\usepackage{upgreek} % Adds in support for greek letters in roman typeset

\newcommand*\patchAmsMathEnvironmentForLineno[1]{%
\expandafter\let\csname old#1\expandafter\endcsname\csname #1\endcsname
\expandafter\let\csname oldend#1\expandafter\endcsname\csname
end#1\endcsname
 \renewenvironment{#1}%
   {\linenomath\csname old#1\endcsname}%
   {\csname oldend#1\endcsname\endlinenomath}%
}
\newcommand*\patchBothAmsMathEnvironmentsForLineno[1]{%
  \patchAmsMathEnvironmentForLineno{#1}%
  \patchAmsMathEnvironmentForLineno{#1*}%
}
\AtBeginDocument{%
\patchBothAmsMathEnvironmentsForLineno{equation}%
\patchBothAmsMathEnvironmentsForLineno{align}%
\patchBothAmsMathEnvironmentsForLineno{flalign}%
\patchBothAmsMathEnvironmentsForLineno{alignat}%
\patchBothAmsMathEnvironmentsForLineno{gather}%
\patchBothAmsMathEnvironmentsForLineno{multline}%
\patchBothAmsMathEnvironmentsForLineno{eqnarray}%
}

\usepackage{hyperxmp}

\usepackage[pdftex,
            pdfauthor={\paperauthors},
            pdftitle={\paperasciititle},
            pdfkeywords={\paperkeywords},
            pdfcopyright={Copyright (C) \papercopyright},
            pdflicenseurl={\paperlicenceurl}]{hyperref}

\usepackage[colorinlistoftodos,textsize=scriptsize]{todonotes}

\usepackage[all]{hypcap} % Internal hyperlinks to floats.

\usepackage{xspace} 
\usepackage{upgreek}

\def\lhcb   {\mbox{LHCb}\xspace}

\def\MagUp {\mbox{\em Mag\kern -0.05em Up}\xspace}

\ifthenelse{\boolean{uprightparticles}}%
{

 \def\Ppi         {\ensuremath{\uppi}\xspace}

 \def\Ppsi        {\ensuremath{\uppsi}\xspace}

 \def\PDelta      {\ensuremath{\Delta}\xspace}                 
 \def\PXi         {\ensuremath{\Xi}\xspace}                 
 \def\PLambda     {\ensuremath{\Lambda}\xspace}                 
 \def\PSigma      {\ensuremath{\Sigma}\xspace}                 
 \def\POmega      {\ensuremath{\Omega}\xspace}                 
 \def\PUpsilon    {\ensuremath{\Upsilon}\xspace}
 \let\oldPi\Pi
 \def\PPi         {\ensuremath{\oldPi}\xspace}

 \def\PB      {\ensuremath{\mathrm{B}}\xspace}                 
                  
 \def\PD      {\ensuremath{\mathrm{D}}\xspace}

 \def\PJ      {\ensuremath{\mathrm{J}}\xspace}                 
 \def\PK      {\ensuremath{\mathrm{K}}\xspace}

 \def\Pb      {\ensuremath{\mathrm{b}}\xspace}                 
 \def\Pc      {\ensuremath{\mathrm{c}}\xspace}

 \def\Pi      {\ensuremath{\mathrm{i}}\xspace}

 \def\Ps      {\ensuremath{\mathrm{s}}\xspace}

 \def\thebaroffset{0.0em}
}
{

 \def\Ppi         {\ensuremath{\pi}\xspace}

 \def\Ppsi        {\ensuremath{\psi}\xspace}                 
                  
 \mathchardef\PDelta="7101
 \mathchardef\PXi="7104
 \mathchardef\PLambda="7103
 \mathchardef\PSigma="7106
 \mathchardef\POmega="710A
 \mathchardef\PUpsilon="7107
 \mathchardef\PPi="7105
                  
 \def\PB      {\ensuremath{B}\xspace}                 
                  
 \def\PD      {\ensuremath{D}\xspace}

 \def\PJ      {\ensuremath{J}\xspace}                 
 \def\PK      {\ensuremath{K}\xspace}

 \def\Pb      {\ensuremath{b}\xspace}                 
 \def\Pc      {\ensuremath{c}\xspace}

 \def\Pi      {\ensuremath{i}\xspace}

 \def\Ps      {\ensuremath{s}\xspace}

 \def\thebaroffset{0.18em}
}
\newcommand{\offsetoverline}[2][\thebaroffset]{\kern #1\overline{\kern -#1 #2}}%

\makeatletter
\ifcase \@ptsize \relax% 10pt
  \newcommand{\miniscule}{\@setfontsize\miniscule{4}{5}}% \tiny: 5/6
\or% 11pt
  \newcommand{\miniscule}{\@setfontsize\miniscule{5}{6}}% \tiny: 6/7
\or% 12pt
  \newcommand{\miniscule}{\@setfontsize\miniscule{5}{6}}% \tiny: 6/7
\fi
\makeatother

\DeclareRobustCommand{\optbar}[1]{\shortstack{{\miniscule (\rule[.5ex]{1.25em}{.18mm})}
  \\ [-.7ex] $#1$}}

\def\squark    {{\ensuremath{\Ps}}\xspace}

\def\cquark    {{\ensuremath{\Pc}}\xspace}
\def\cquarkbar {{\ensuremath{\overline \cquark}}\xspace}
\def\ccbar     {{\ensuremath{\cquark\cquarkbar}}\xspace}
\def\bquark    {{\ensuremath{\Pb}}\xspace}

\def\pion   {{\ensuremath{\Ppi}}\xspace}

\def\pip    {{\ensuremath{\pion^+}}\xspace}
\def\pim    {{\ensuremath{\pion^-}}\xspace}

\def\kaon    {{\ensuremath{\PK}}\xspace}
\def\KorKbar {\kern \thebaroffset\optbar{\kern -\thebaroffset \PK}{}\xspace}

\def\Kp      {{\ensuremath{\kaon^+}}\xspace}
\def\Km      {{\ensuremath{\kaon^-}}\xspace}

\def\D       {{\ensuremath{\PD}}\xspace}

\def\DorDbar {\kern \thebaroffset\optbar{\kern -\thebaroffset \PD}\xspace}

\def\Dp      {{\ensuremath{\D^+}}\xspace}
\def\Dm      {{\ensuremath{\D^-}}\xspace}

\def\DpDm    {\ensuremath{\Dp {\kern -0.16em \Dm}}\xspace}

\def\B       {{\ensuremath{\PB}}\xspace}
\def\Bbar    {{\ensuremath{\offsetoverline{\PB}}}\xspace}

\def\BorBbar {\kern \thebaroffset\optbar{\kern -\thebaroffset \PB}\xspace}

\def\Bd      {{\ensuremath{\B^0}}\xspace}

\def\BdorBdbar {\kern \thebaroffset\optbar{\kern -\thebaroffset \Bd}\xspace}

\def\Bs      {{\ensuremath{\B^0_\squark}}\xspace}
\def\Bsb     {{\ensuremath{\Bbar{}^0_\squark}}\xspace}
\def\BsorBsbar {\kern \thebaroffset\optbar{\kern -\thebaroffset \Bs}\xspace}

\def\jpsi     {{\ensuremath{{\PJ\mskip -3mu/\mskip -2mu\Ppsi}}}\xspace}

\def\Y#1S{\ensuremath{\PUpsilon{(#1S)}}\xspace}

\def\Lz          {{\ensuremath{\PLambda}}\xspace}

\def\LorLbar     {\kern \thebaroffset\optbar{\kern -\thebaroffset \PLambda}\xspace}

\def\Lb           {{\ensuremath{\Lz^0_\bquark}}\xspace}

\def\to                 {\ensuremath{\rightarrow}\xspace}

\def\CP                {{\ensuremath{C\!P}}\xspace}

\newcommand{\phis}{{\ensuremath{\phi_{\squark}}}\xspace}
\newcommand{\betas}{{\ensuremath{\beta_{\squark}}}\xspace}

\def\AT#1     {\ensuremath{A_{\mathrm{T}}^{#1}}\xspace}           % 2

\def\C#1      {\ensuremath{\mathcal{C}_{#1}}\xspace}                       % 9
\def\Cp#1     {\ensuremath{\mathcal{C}_{#1}^{'}}\xspace}                    % 7
\def\Ceff#1   {\ensuremath{\mathcal{C}_{#1}^{\mathrm{(eff)}}}\xspace}        % 9  
\def\Cpeff#1  {\ensuremath{\mathcal{C}_{#1}^{'\mathrm{(eff)}}}\xspace}       % 7
\def\Ope#1    {\ensuremath{\mathcal{O}_{#1}}\xspace}                       % 2
\def\Opep#1   {\ensuremath{\mathcal{O}_{#1}^{'}}\xspace}                    % 7

\newcommand{\aunit}[1]{\ensuremath{\text{\,#1}}}       
\newcommand{\tev}{\aunit{Te\kern -0.1em V}\xspace}
\newcommand{\gev}{\aunit{Ge\kern -0.1em V}\xspace}
\newcommand{\mev}{\aunit{Me\kern -0.1em V}\xspace}
\newcommand{\kev}{\aunit{ke\kern -0.1em V}\xspace}
\newcommand{\ev}{\aunit{e\kern -0.1em V}\xspace}
 
\newcommand{\mevc}{\ensuremath{\aunit{Me\kern -0.1em V\!/}c}\xspace}
\newcommand{\gevc}{\ensuremath{\aunit{Ge\kern -0.1em V\!/}c}\xspace}
\newcommand{\mevcc}{\ensuremath{\aunit{Me\kern -0.1em V\!/}c^2}\xspace}
\newcommand{\gevcc}{\ensuremath{\aunit{Ge\kern -0.1em V\!/}c^2}\xspace}
\def\fb   {\ensuremath{\aunit{fb}}\xspace}
\def\invfb   {\ensuremath{\fb^{-1}}\xspace}

\def\ps   {\ensuremath{\aunit{ps}}\xspace}

\def\invps{\ensuremath{\ps^{-1}}\xspace}

\def\gsim{{~\raise.15em\hbox{$>$}\kern-.85em
          \lower.35em\hbox{$\sim$}~}\xspace}
\def\lsim{{~\raise.15em\hbox{$<$}\kern-.85em
          \lower.35em\hbox{$\sim$}~}\xspace}

\def\sPlot{\mbox{\em sPlot}\xspace}

\def\rad{\aunit{rad}\xspace}

\def\tell1  {TELL1\xspace}
\def\ukl1   {UKL1\xspace}

\newcommand{\lhcborcid}[1]{\href{https://orcid.org/#1}{\hspace*{0.1em}\raisebox{-0.45ex}{\includegraphics[width=1em]{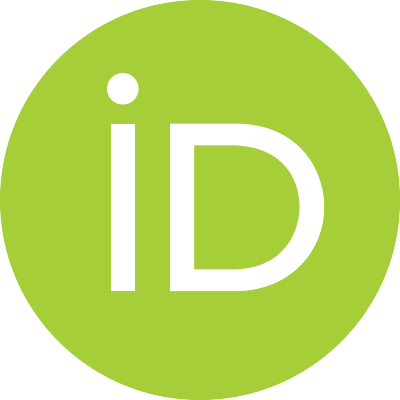}}}}

\newcommand{\Jpsi}{\ensuremath{J/\psi}\xspace}

\newcommand{\BsJpsiKK}{\ensuremath{B^0_s\to J/\psi K^+K^-}\xspace}

\newcommand{\figref}[1]{Fig.~\ref{#1}}
\newcommand{\tabref}[1]{Table~\ref{#1}}

\newcommand{\lamfbar}[1][\text{f}]{{\kern 0.06em \overline{\kern -0.06em \lambda \kern -0.03em}\kern 0.03em}_{\overline{\text{#1}}}}

\newcommand{\AS}[1][]{A_\text{S}#1}
\newcommand{\AAv}[1][\Ai]{#1^{\text{CP}}}

\newcommand{\magSAv}[1][]{|\AAv[\AS]|}
\newcommand{\magSAvSq}[1][]{\magSAv^2}

\newcommand{\definemath}[2]{\newcommand{#1}{\ensuremath{#2}\xspace}}
\definemath{\phisRunIICV}{-0.039}
\definemath{\phisRunIIStatErr}{0.022}
\definemath{\phisRunIISystErr}{0.006}

\definemath{\DGsRunIICV}{0.0845}
\definemath{\DGsRunIIStatErr}{0.0044}
\definemath{\DGsRunIISystErr}{0.0024}

\definemath{\GsGdRunIICV}{-0.0056}
\definemath{\GsGdRunIIStatErrUpper}{0.0013}
\definemath{\GsGdRunIIStatErrLower}{0.0015}
\definemath{\GsGdRunIISystErr}{0.0014}

\definemath{\lambRunIICV}{1.001}
\definemath{\lambRunIIStatErr}{0.011}
\definemath{\lambRunIISystErr}{0.005}

\definemath{\DmsRunIICV}{17.743}
\definemath{\DmsRunIIStatErr}{0.033}
\definemath{\DmsRunIISystErr}{0.009}

\definemath{\AperpSqRunIICV}{0.2463}
\definemath{\AperpSqRunIIStatErr}{0.0023}
\definemath{\AperpSqRunIISystErr}{0.0024}

\definemath{\AzeroSqRunIICV}{0.5179}
\definemath{\AzeroSqRunIIStatErr}{0.0017}
\definemath{\AzeroSqRunIISystErr}{0.0032}

\definemath{\DperpRunIICV}{2.903}
\definemath{\DperpRunIIStatErrUpper}{0.075}
\definemath{\DperpRunIIStatErrLower}{0.074}
\definemath{\DperpRunIISystErr}{0.048}

\definemath{\DparRunIICV}{3.146}
\definemath{\DparRunIIStatErr}{0.061}
\definemath{\DparRunIISystErr}{0.052}

\definemath{\ASISqRunIICV}{0.472}
\definemath{\ASISqRunIIStatErr}{0.024}
\definemath{\ASISqRunIISystErr}{0.027}

\definemath{\ASIISqRunIICV}{0.042}
\definemath{\ASIISqRunIIStatErrUpper}{0.0048}
\definemath{\ASIISqRunIIStatErrLower}{0.0046}
\definemath{\ASIISqRunIISystErr}{0.010}

\definemath{\ASIIISqRunIICV}{0.0029}
\definemath{\ASIIISqRunIIStatErrUpper}{0.0013}
\definemath{\ASIIISqRunIIStatErrLower}{0.0009}
\definemath{\ASIIISqRunIISystErr}{0.023}

\definemath{\ASIVSqRunIICV}{0.0037}
\definemath{\ASIVSqRunIIStatErrUpper}{0.0025}
\definemath{\ASIVSqRunIIStatErrLower}{0.0019}
\definemath{\ASIVSqRunIISystErr}{0.032}

\definemath{\ASVSqRunIICV}{0.0508}
\definemath{\ASVSqRunIIStatErrUpper}{0.0070}
\definemath{\ASVSqRunIIStatErrLower}{0.0068}
\definemath{\ASVSqRunIISystErr}{0.027}

\definemath{\ASVISqRunIICV}{0.151}
\definemath{\ASVISqRunIIStatErr}{0.011}
\definemath{\ASVISqRunIISystErr}{0.051}

\definemath{\DSISqRunIICV}{2.05}
\definemath{\DSISqRunIIStatErrUpper}{0.12}
\definemath{\DSISqRunIIStatErrLower}{0.14}
\definemath{\DSISqRunIISystErr}{0.19}

\definemath{\DSIISqRunIICV}{1.62}
\definemath{\DSIISqRunIIStatErrUpper}{0.19}
\definemath{\DSIISqRunIIStatErrLower}{0.19}
\definemath{\DSIISqRunIISystErr}{0.41}

\definemath{\DSIIISqRunIICV}{1.16}
\definemath{\DSIIISqRunIIStatErrUpper}{0.37}
\definemath{\DSIIISqRunIIStatErrLower}{0.29}
\definemath{\DSIIISqRunIISystErr}{0.19}

\definemath{\DSIVSqRunIICV}{-0.15}
\definemath{\DSIVSqRunIIStatErrUpper}{0.12}
\definemath{\DSIVSqRunIIStatErrLower}{0.15}
\definemath{\DSIVSqRunIISystErr}{0.31}
\definemath{\DSVSqRunIICV}{-0.637}
\definemath{\DSVSqRunIIStatErrUpper}{0.068}
\definemath{\DSVSqRunIIStatErrLower}{0.076}
\definemath{\DSVSqRunIISystErr}{0.17}
\definemath{\DSVISqRunIICV}{-1.013}
\definemath{\DSVISqRunIIStatErrUpper}{0.074}
\definemath{\DSVISqRunIIStatErrLower}{0.083}
\definemath{\DSVISqRunIISystErr}{0.07}
\definemath{\phisAvgJpsiKKCV}{-0.044}
\definemath{\phisAvgJpsiKKErr}{0.020}
\definemath{\lambAvgJpsiKKCV}{0.990}
\definemath{\lambAvgJpsiKKErr}{0.010}
\definemath{\phisAvgAllCV}{-0.031}
\definemath{\phisAvgAllErr}{0.018}
\usepackage{cite} % Allows for ranges in citations
\usepackage{mciteplus}
\mciteErrorOnUnknownfalse
\usepackage{multirow}
\usepackage{listings}

\usepackage{makecell}
\usepackage{booktabs}
\usepackage{rotating}

\definecolor{pleasantgreen}{rgb}{0.3,0.7,0.3}

\definemath{\BsYieldFifteen}{16\,181}
\definemath{\BsYieldFifteenErr}{135}

\definemath{\BsYieldSixteen}{103\,319}
\definemath{\BsYieldSixteenErr}{342}

\definemath{\BsYieldSeventeen}{105\,465}
\definemath{\BsYieldSeventeenErr}{343}

\definemath{\BsYieldEighteen}{123\,870}
\definemath{\BsYieldEighteenErr}{476}

\usepackage{verbatim}

\usepackage{longtable} % only for template; not usually to be used in PAPERs
\usepackage{tablefootnote}
\begin{document}

%%%%%%%%%%%%%%%%%%%%%%%%%
%%%%% Title     %%%%%%%%%
%%%%%%%%%%%%%%%%%%%%%%%%%
\renewcommand{\thefootnote}{\fnsymbol{footnote}}
\setcounter{footnote}{1}

% %%%%%%% CHOOSE TITLE PAGE--------
%\onecolumn
%\input{title-LHCb-INT}
%\input{title-LHCb-ANA}
%\input{title-LHCb-CONF}
%\input{title-LHCb-FIGURE}
%TC:ignore
% ===============================================================================
% Purpose: LHCb-PAPER journal paper title page template
% Author:
% Created on: 2010-09-25
% ===============================================================================

%%%%%%%%%%%%%%%%%%%%%%%%%
%%%%%  TITLE PAGE  %%%%%%
%%%%%%%%%%%%%%%%%%%%%%%%%
\begin{titlepage}
\pagenumbering{roman}

% Header ---------------------------------------------------
\vspace*{-1.5cm}
\centerline{\large EUROPEAN ORGANIZATION FOR NUCLEAR RESEARCH (CERN)}
\vspace*{1.5cm}
\noindent
\begin{tabular*}{\linewidth}{lc@{\extracolsep{\fill}}r@{\extracolsep{0pt}}}
\ifthenelse{\boolean{pdflatex}}% Logo format choice
{\vspace*{-1.5cm}\mbox{\!\!\!\includegraphics[width=.14\textwidth]{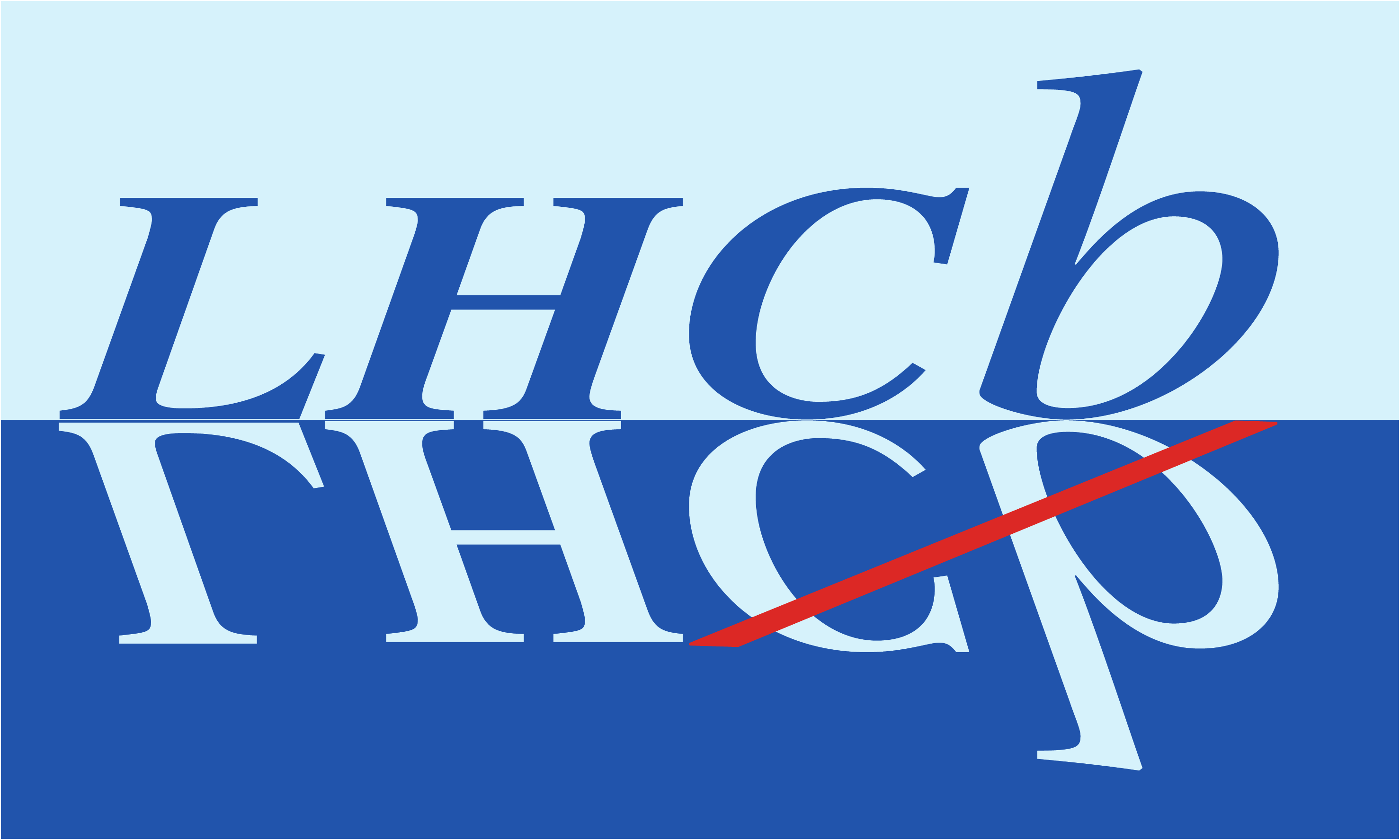}} & &}%
{\vspace*{-1.2cm}\mbox{\!\!\!\includegraphics[width=.12\textwidth]{lhcb-logo.eps}} & &}%
\\
 & & CERN-EP-2023-145\\  % ID
 & & LHCb-PAPER-2023-016 \\  % ID
 & & February 21, 2024 \\ % Date - Can also hardwire e.g.: 23 March 2010
 & & \\
% not in paper \hline
\end{tabular*}

\vspace*{2.0cm}

% Title --------------------------------------------------
{\normalfont\bfseries\boldmath\huge
\begin{center}
% DO NOT EDIT HERE. Instead edit macro in main.tex to keep metadata correct
  \papertitle
\end{center}
}

\vspace*{0.5cm}

% Authors -------------------------------------------------
\begin{center}
%In the footnote, replace 'paper' by 'Letter' in case of submission to PRL or PLB
% Edit macro in main.tex to keep metadata correct
\paperauthors\footnote{Full author list given at the end of the Letter.}
\end{center}

%\vspace{\fill}
\vspace{0.5cm}

% Abstract -----------------------------------------------
\begin{abstract}
The decay-time-dependent \CP asymmetry in $\Bs\to\jpsi(\to\mu^+\mu^-) \Kp \Km$ decays is measured using proton-proton collision data, corresponding to an integrated luminosity of
    $6\invfb$, collected with the LHCb detector at a center-of-mass energy of 13~\tev. Using a sample of approximately 349\,000 \Bs signal decays with an invariant $\Kp \Km$ mass in the vicinity of the $\phi(1020)$ resonance, the \CP-violating
    phase $\phi_s$ is measured, along with the difference in decay widths of the
    light and heavy mass eigenstates of the $\Bs$-$\Bsb$ system, $\Delta\Gamma_s$, and the difference of the average $\Bs$ and $\Bd$ meson decay widths, $\Gamma_s-\Gamma_d$. 
    The values obtained are
    \mbox{$\phi_s = \phisRunIICV\pm\phisRunIIStatErr\pm\phisRunIISystErr\rad$}, \mbox{$\Delta\Gamma_s = \DGsRunIICV \pm \DGsRunIIStatErr \pm \DGsRunIISystErr \invps$},
    and \mbox{$\Gamma_s-\Gamma_d = \GsGdRunIICV^{\:+\:\GsGdRunIIStatErrUpper}_{\:-\:\GsGdRunIIStatErrLower} \pm \GsGdRunIISystErr \invps$},
    where the first
    uncertainty is statistical and the second systematic. These are the most
    precise single measurements to date and are consistent
    with expectations based on the Standard Model and with the previous LHCb 
    analyses of this decay. 
    These results are combined with previous independent LHCb measurements. The phase $\phi_s$ is also measured independently for each polarization state of the $\Kp\Km$ system and shows no evidence for polarization dependence.
\end{abstract}

\vspace*{0.5cm}

\begin{center}
Published in Phys. Rev. Lett. 132 (2024) 051802
\end{center}

\vspace{\fill}

{\footnotesize
% Edit macro in main.tex to keep metadata correct
\centerline{\copyright~\papercopyright. \href{\paperlicenceurl}{\paperlicence}.}}
\vspace*{2mm}

\end{titlepage}

%%%%%%%%%%%%%%%%%%%%%%%%%%%%%%%%
%%%%%  EOD OF TITLE PAGE  %%%%%%
%%%%%%%%%%%%%%%%%%%%%%%%%%%%%%%%

%  empty page follows the title page ----
\newpage
\setcounter{page}{2}
\mbox{~}

\cleardoublepage

%TC:endignore

%\twocolumn
% %%%%%%%%%%%%% ---------

\renewcommand{\thefootnote}{\arabic{footnote}}
\setcounter{footnote}{0}

%%%%%%%%%%%%%%%%%%%%%%%%%%%%%%%%
%%%%%  Table of Content   %%%%%%
%%%%%%%%%%%%%%%%%%%%%%%%%%%%%%%%
%%%% Uncomment if desired
%\tableofcontents
\cleardoublepage

%%%%%%%%%%%%%%%%%%%%%%%%%
%%%%% Main text %%%%%%%%%
%%%%%%%%%%%%%%%%%%%%%%%%%

\pagestyle{plain} % restore page numbers for the main text
\setcounter{page}{1}
\pagenumbering{arabic}

%% Uncomment during review phase. 
%% Comment before a final submission.
%\linenumbers

%% This is the main body
%% It is useful to have a single file so comemnts are not missed in overleaf.
The interference between \Bs-mixing and -decay amplitudes
to \CP eigenstates with a \ccbar resonance in the final state 
gives rise to a measurable \CP-violating phase \phis, which is particularly sensitive to physics beyond the standard model (SM). 
 In the SM, neglecting subleading loop contributions in $b\to c\overline{c}s$ transitions, \phis is predicted to be equal to $-2\betas$, where $\betas\equiv\arg\left[
 - (V_{ts} V_{tb}^*) / (V_{cs} V_{cb}^*)\right]$ and $V_{ij}$ are elements of the Cabibbo-Kobayashi-Maskawa (CKM) quark-flavor-mixing matrix~\cite{Kobayashi:1973fv,*Cabibbo:1963yz}. Global fits to experimental data, under the assumption of the CKM paradigm, give
$-2\beta_s = -0.0368 ^{+0.0009}_{-0.0006}~\rad$ ~\cite{CKMfitter2015}. This precise indirect determination makes the
measurement of \phis\ an excellent probe for physics beyond the SM, especially for models contributing to $\Bs-\Bsb$ mixing~\cite{Buras:2009if,Dutta2008}.
Several experiments have measured  \phis, the \Bs decay width, and decay-width difference, $
\Gamma_s$ and $\Delta\Gamma_{s}$, respectively, in \Bs decays via $b\to c\bar{c}s$ transitions~\cite{Aaltonen:2012ie,Abazov:2011ry,Aad:2014cqa,Aad:2016tdj,Aad:2020jfw,Khachatryan:2015nza,Sirunyan:2020vke,LHCb-PAPER-2014-059,LHCb-PAPER-2019-013,LHCb-PAPER-2020-042,LHCb-PAPER-2017-008,LHCb-PAPER-2014-019,LHCb-PAPER-2019-003,LHCb-PAPER-2016-027,LHCb-PAPER-2014-051}. The measurements lead to the current world average of $\phi_s^{c\bar{c}s}=-0.049\pm 0.019~\rad$~\cite{PDG2022}, which is dominated by the LHCb results in $B_s^0\to\jpsi h^+h^-$ ($h=K,\pi$) decays using 5~\invfb of data collected at center-of-mass energies of $\sqrt{s}=$7, 8 and 13\tev~\cite{LHCb-PAPER-2019-013}.
 This Letter reports an update of the LHCb measurements in \BsJpsiKK decays~\cite{Charge-Conj-footnote} using the full Run~2 data taken in 2015--2018, corresponding to an integrated luminosity of 6~\invfb at $\sqrt{s}=$13\tev. The results supersede those in Ref.~\cite{LHCb-PAPER-2019-013} and are combined with the LHCb result using 3~\invfb of Run 1 data~\cite{LHCb-PAPER-2014-059}.
Apart from the increased data size, this analysis benefits from improvements in the calibration procedures of the particle identification (PID), flavor-tagging algorithms, and the \Bs decay-time resolution model.

The \lhcb detector is a single-arm forward
spectrometer covering the \mbox{pseudorapidity} range $2<\eta <5$, described in detail in Refs~\cite{LHCb-DP-2008-001,LHCb-DP-2014-002}. 
Simulated events produced with the software described in Refs.~\cite{Sjostrand:2007gs,*Sjostrand:2006za,LHCb-PROC-2010-056,Lange:2001uf,davidson2015photos}
are used to model the effects of the detector acceptance, resolution, and selection requirements.
Dedicated simulations are produced for each year of data taking corresponding to the 
relevant detector and accelerator conditions.
High-purity data samples of charm hadron, charmonia, and beauty hadron decays
are used to calibrate the simulated single-particle reconstruction and PID efficiencies.
The trigger, which performs the online data selection~\cite{LHCb-DP-2012-004}, consists of a hardware stage, based on the high transverse momentum, $p_{T}$, and signatures from the calorimeter and muon
systems, followed by a software stage with full event
reconstruction. 
Two software trigger algorithms, exploiting the muon kinematics and two-muon vertex information, are used to select candidates. The first requires a vertex with a large impact-parameter significance from all primary vertexes and, therefore, introduces a nontrivial efficiency dependence on the $B_{s}^{0}$ decay time. The second has almost uniform decay-time efficiency. Consequently, candidates selected by the two triggers are studied in separate categories.

The selection of $B_s^0 \to J/\psi (\to \mu^+ \mu^-) K^+ K^-$ candidates, with $K^+K^-$ invariant masses in the range [990, 1050] $\mevcc$, follows the same strategy used in the previous Run~2 measurement~\cite{LHCb-PAPER-2019-013}. In order to take into account different data-taking and calibration conditions for optimal event selection, a gradient-boosted decision tree (BDT) classifier applied in the selection is trained separately for each year between 2016 and 2018, with the result for 2016 applied to the 2015 dataset due to its limited size. The BDT selection improves the signal-to-background ratio by about a factor of 50.
The peaking backgrounds due to pion and proton misidentification in \Bd and \Lb decays are significantly reduced with stringent PID and mass requirements. The remaining peaking background from 
\mbox{$\Lambda_b^0 \rightarrow \jpsi p \Km$} decays is subtracted statistically through the injection of simulated events into the data with a negative sum of weights equal to the expected number of 4700 \Lb candidates.

\begin{figure}[tb]
\centering
\includegraphics[width=.48\textwidth]{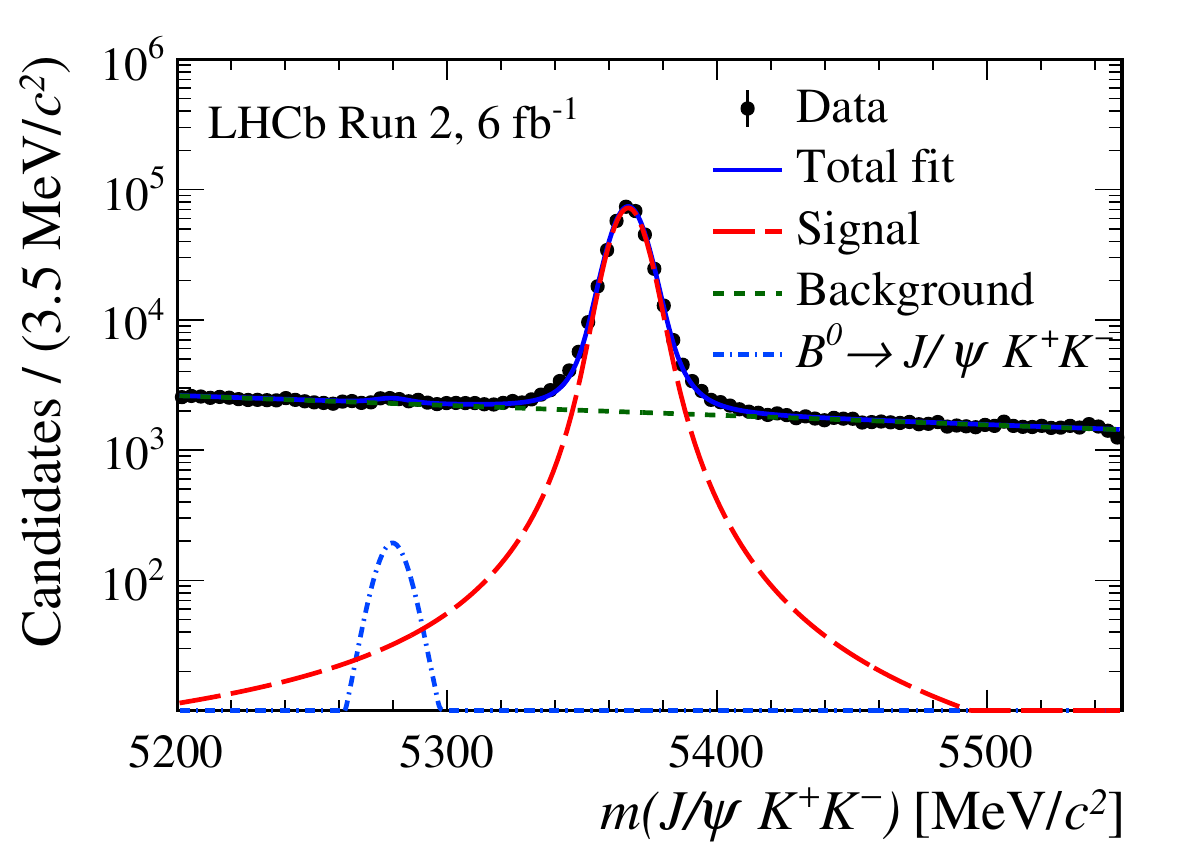}
\caption{\small
Distribution of $m(\jpsi\Kp\Km)$ for the full data sample and projection of the maximum likelihood fit.
}
\label{fig:mass_plots_fit_signal_region}
\end{figure}

Selected $B_s^0\to\jpsi\Kp\Km$ candidates in the mass range [5200, 5550]~$\mevcc$ 
are subsequently retained for analysis. 
The data sample is divided into 48 independent subsamples, corresponding to six $m(\Kp\Km)$ bins with boundaries at 990, 1008, 1016, 1020, 1024, 1032 and 1050~$\mevcc$, 
two trigger categories, and four years of data taking. 
The invariant mass of selected \Bs candidates, $m(\jpsi\Kp\Km)$, and the per-candidate mass uncertainty, $\sigma_m$, are calculated by constraining the $\jpsi$ mass to the world average~\cite{PDG2022} and requiring the \Bs candidate momentum to point back to the corresponding primary vertex. Using $m(\jpsi\Kp\Km)$ as the discriminating variable, a signal weight is assigned to each candidate with the \sPlot\ method~\cite{splot,sFit, Dembinski_2022}, using an extended maximum-likelihood fit, shown in Fig.~\ref{fig:mass_plots_fit_signal_region}.
The signal shape is described by a double-sided Crystal Ball (CB) function ~\cite{Skwarnicki:1986xj},
 whose width is parametrized as a function of $\sigma_m$, using a second-order polynomial. This parametrization accounts for the correlation between $m(\jpsi\Kp\Km)$ and the helicity angle $\cos\theta_\mu$, which is due to the dependence of the $m(\jpsi\Kp\Km)$ resolution, characterized by $\sigma_m$, on the $p_{T}$ of the muons. Since the muon $p_T$ depends on $\cos\theta_\mu$, $\sigma_m$ is found to represent a good proxy for $\cos\theta_\mu$. The parameters that describe the tail of the CB function are fixed to those obtained from simulation.
The background from $B^0\to \jpsi\Kp\Km$ decays is modeled with the same CB function as the signal, sharing all shape parameters except for the mean of the distribution. The difference between the means of the signal and \Bd components is fixed to its world average~\cite{PDG2022}. The background due to random combinations of tracks is modeled with an exponential function. The peaking background from $B^0\to\jpsi \Kp\pim$ decays is estimated to be negligible.
The \BsJpsiKK signal yields are
$16\,181 \pm 135$,
$103\,319 \pm 342$,
$105\,465 \pm 343$ and
$123\,870 \pm 476$ for the 2015, 2016, 2017 and 2018 datasets, respectively.

The measurement of \phis in \BsJpsiKK decays requires the \CP-even and \CP-odd decay-amplitude components to be disentangled, depending upon the relative orbital angular momentum between the $\jpsi$ candidate and the kaon pair. 
A weighted simultaneous fit to the distributions of decay time and decay angles ($\cos\theta_K, \cos\theta_\mu, \phi_h$) in the helicity basis, as described in Ref.~\cite{LHCb-PAPER-2019-013}, is performed for the 48 independent subsamples, to determine the physics parameters. 
These parameters are \phis; $|\lambda|$; $\Gamma_s-\Gamma_d$; $\Delta\Gamma_s$; the \Bs mass difference $\Delta m_s$;  and the polarization amplitudes $A_k=|A_k|e^{-i\delta_k}$, where the indices $k\in {\{0,\parallel,\perp, S\}}$ refer to the different polarization states of the $\Kp\Km$ system. The sum $|A_\parallel|^2+|A_0|^2+|A_\perp|^2$ equals unity, and $\delta_0$ is zero, by convention. The parameter $\lambda$ is defined as $\eta_k(q/p)(\bar{A}_{k}/A_{k})$, 
where $p = \left<B_s^0|B_{\rm L}\right>$ and $q = \left<\bar{B}_s^0|B_{\rm L}\right>$ describe the relation between mass
and flavor eigenstates and $\eta_{k}$ is the \CP eigenvalue of the polarization state $k$. 
 
The probability density function (PDF) for the signal in each subsample accounts for the decay-time resolution, the decay-time and angular efficiencies, and the flavor tagging. It considers {\it P}- and {\it S}-wave components of the kaon pair from $\phi(1020)$ and $f_0(980)$ decays, while the {\it D}-wave component is neglected~\cite{LHCb-PAPER-2012-040,LHCb-PAPER-2017-008}. The interference of {\it P} and {\it S} waves includes an effective coupling factor $C_{\rm SP}$, determined in each $m(\Kp\Km)$ bin through integration of the mass line shape interference term. 
The line shape of the $\phi(1020)$ resonance~\cite{LHCb-PAPER-2012-040,LHCb-PAPER-2017-008} is modeled as a relativistic Breit-Wigner distribution, while the $f_0(980)$ resonance is modeled as a Flatt\'{e} amplitude with parameters from Ref.~\cite{LHCb-PAPER-2013-069}. The effect of mass resolution is also accounted for. The computed values of $C_{\rm SP}$ are $0.8458\pm0.0018$, $0.8673\pm0.0004$, $0.8127\pm0.0012$, $0.8558\pm0.0010$, $0.9359\pm0.0004$ and $0.9735\pm0.0001$ from the lowest to the highest $m(\Kp\Km)$ bin.
The value of $\Gamma_d$
is fixed to its world average~\cite{HFLAV21}.
All physics parameters are left unconstrained in the fit and are shared across the subsamples,
except for the {\it S}-wave fraction and the phase difference
$\delta_{\rm S}-\delta_{\perp}$, which are independent parameters for each $m(\Kp\Km)$ bin.

The experimental decay-time resolution is accounted for by convolving the signal PDF with a Gaussian resolution function with the per-candidate decay-time uncertainty as the width. The per-candidate decay-time uncertainty is calibrated to represent the effective resolution, which is determined from a control sample of promptly decaying $\jpsi$ candidates combined with two kaons selected similarly as the signal except for the decay time and flight distance requirements. The candidates with negative reconstructed decay times, arising from purely detector resolution effects, are used for the calibration. The possible contamination from nonprompt decays is estimated to be around 1\%--2\% and has a negligible effect on the calibration model. The average resolution for the signal candidates is determined to be around 42~fs.

The prompt sample used in the decay-time resolution calibration has a nonzero mean decay time due to residual detector misalignment. This bias, which depends on the kinematics, is corrected for in the analysis. In the misaligned simulated samples, a small bias remains after the correction and is assigned as a systematic uncertainty.

The reconstruction and selection produce a nonuniform efficiency as a function of the decay time and angles of the \Bs decays. 
The angular and decay-time efficiencies are assumed to factorize and are evaluated separately for different years of data taking and for the two trigger categories.  The three-dimensional angular efficiency correction is introduced through normalization weights in the PDF describing the signal decays in the time-dependent angular fit. The efficiency is determined from simulated signal events subjected to the same selection criteria as data. The simulated sample is corrected by an iterative procedure using data~\cite{LHCb-PAPER-2019-013}.

The decay-time efficiency is determined using a data-driven method with a reference channel $B^0 \to \jpsi K^{\ast0}(\to\Kp \pim)$ that is topologically similar to the signal channel. The decay-time efficiency is modeled by a cubic spline function, determined from the decay-time distribution of selected candidates divided by the expected distribution for the case of perfect acceptance. The latter is modeled by an exponential distribution with the \Bd lifetime~\cite{PDG2022}, convolved with a Gaussian resolution function with a width of 42~fs. Simulated \Bd and \Bs events are used to determine and apply corrections at the level of 3\% to account for kinematic differences between \Bd and \Bs decays. 
The background-subtracted $B^0 \to J/\psi K^{*0}$ candidates are selected using the same strategy as in Ref.\cite{LHCb-PAPER-2019-013}, with an additional requirement on the helicity angle $\cos\theta_{K}<0$, to avoid a large difference between signal and control samples, since pions with $\cos\theta_{K}>0$ tend to have extremely low momenta. 
The decay-time efficiency is validated by replacing \Bs samples with $B^+$ and \Bd samples where the decay widths are measured to be consistent with their corresponding world averages~\cite{PDG2022}.

The flavor of the \Bs meson at production is inferred using two independent classes of flavor-tagging algorithms, the opposite-side (OS) tagger~\cite{LHCb-PAPER-2011-027} and same-side (SS) tagger~\cite{LHCb-PAPER-2015-056}, which exploit specific accompanying $B$ meson decays and signal fragmentation information, respectively.   
Each method yields a tagging decision $Q$, with an estimated mistag probability $\kappa$, for each \Bs meson, where $Q =+1,\; -1$, or $0$, if the meson is classified as \Bs, \Bsb or untagged, respectively. To obtain the correct mistag probability $\omega$, each algorithm is calibrated using a linear function following the same strategy as Ref.~\cite{LHCb-PAPER-2019-013}.
 The calibration of the OS mistag probability uses \mbox{$B^+\to\jpsi\Kp$} decays, for which the value of $\omega$ in an interval of $\kappa$ can be obtained from the number of correct and wrong decisions. 
The calibration of the SS mistag probability uses flavor-specific $B_s^0\to D_s^-\pi^+$ decays, for which the value of $\omega$ in an interval of $\kappa$ is estimated by fitting the decay-time distribution. The decay-time acceptance is modeled with a cubic spline~\cite{LHCb-PAPER-2021-005}. 
The effective tagging power is given by the product of the tagging efficiency ($\epsilon_{\rm tag}$) and the wrong-tag dilution squared, $\epsilon_{\rm tag}\times(1 - 2\omega)^2 $. 
The combined tagging powers of the OS and SS taggers are $(4.18\pm0.15)$\%, $(4.22\pm0.16)$\%, $(4.36\pm0.16)$\% for 2015--2016, 2017, and 2018, respectively, where the statistical and systematic uncertainties are combined. A novel inclusive flavor-tagging algorithm~\cite{IFTYandex}, which uses track information from the full event, is applied as an alternative method to cross-check the OS and SS combined method and provides compatible results for $\phi_{s}$ with similar precision.

\begin{figure}[ptb]
        \centering
       \includegraphics[width=0.48\textwidth]{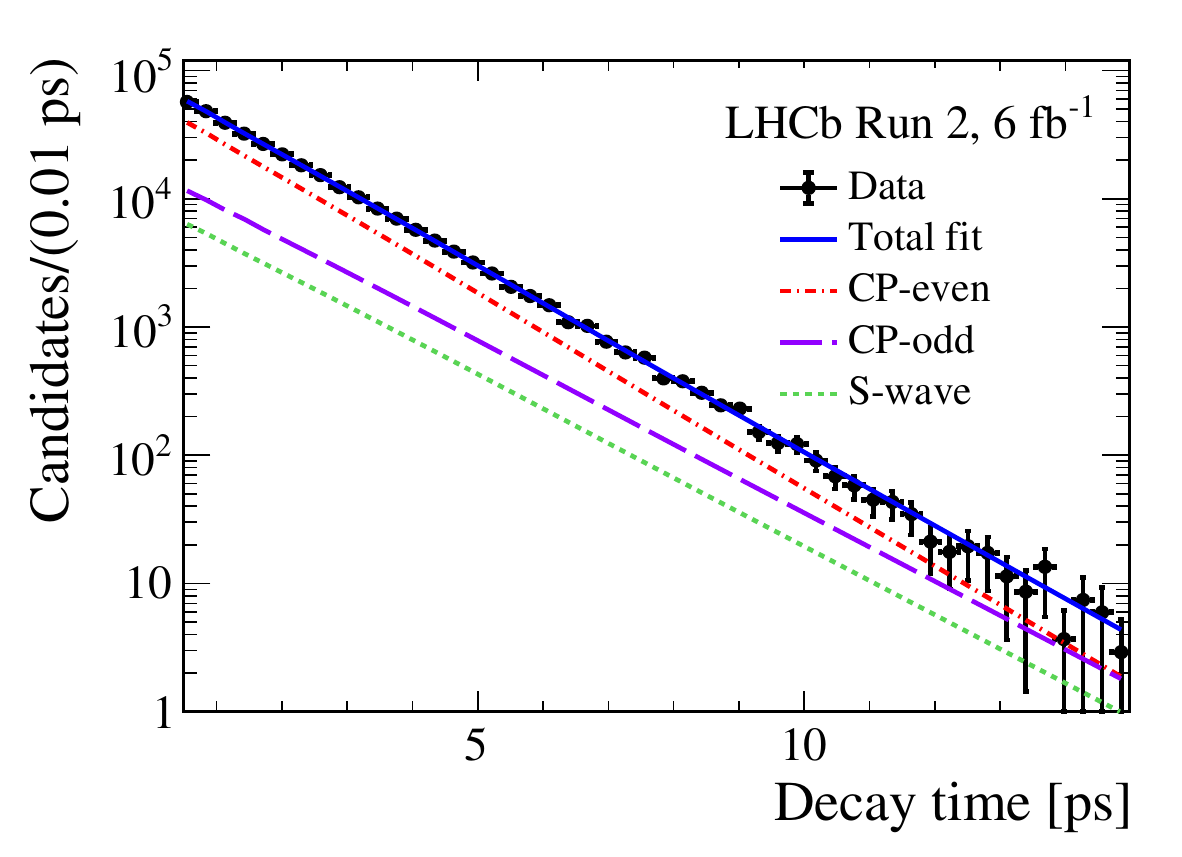}
       \includegraphics[width=0.48\textwidth]{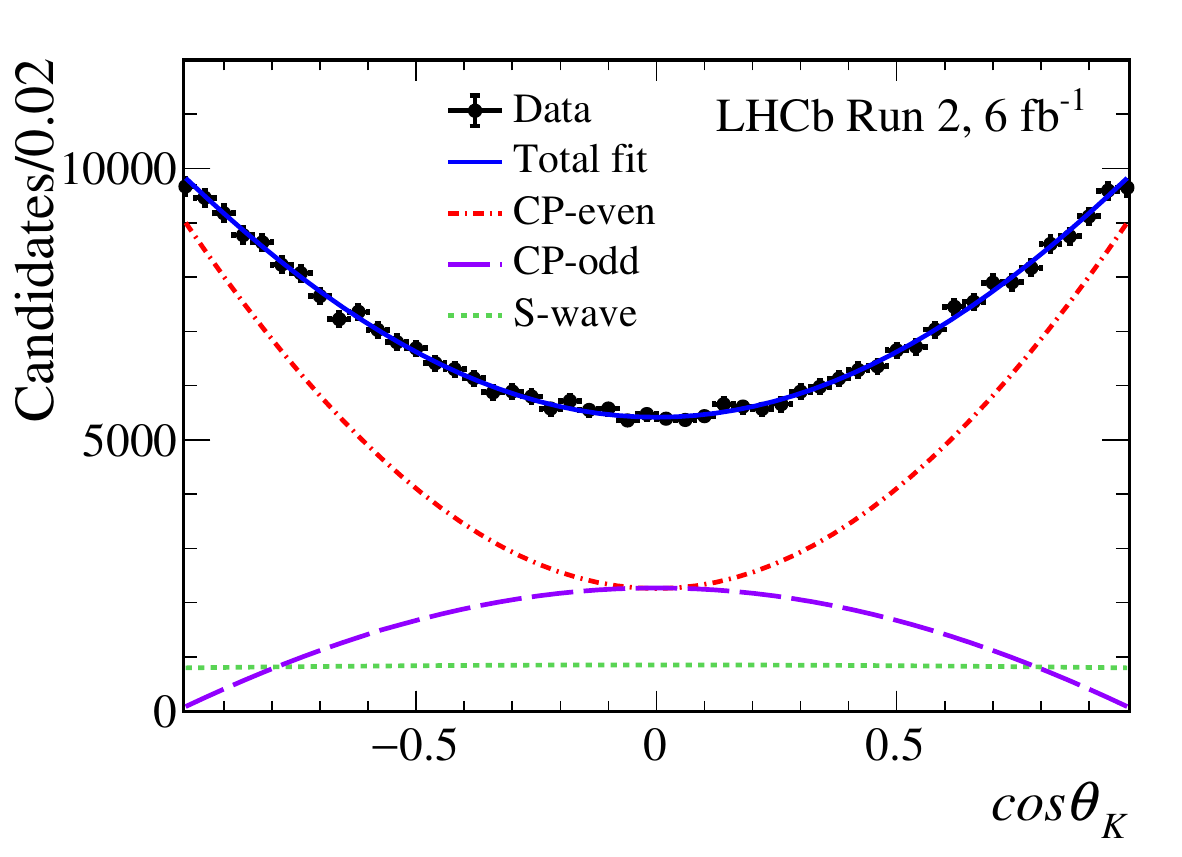}   \\ 
    \includegraphics[width=0.48\textwidth]{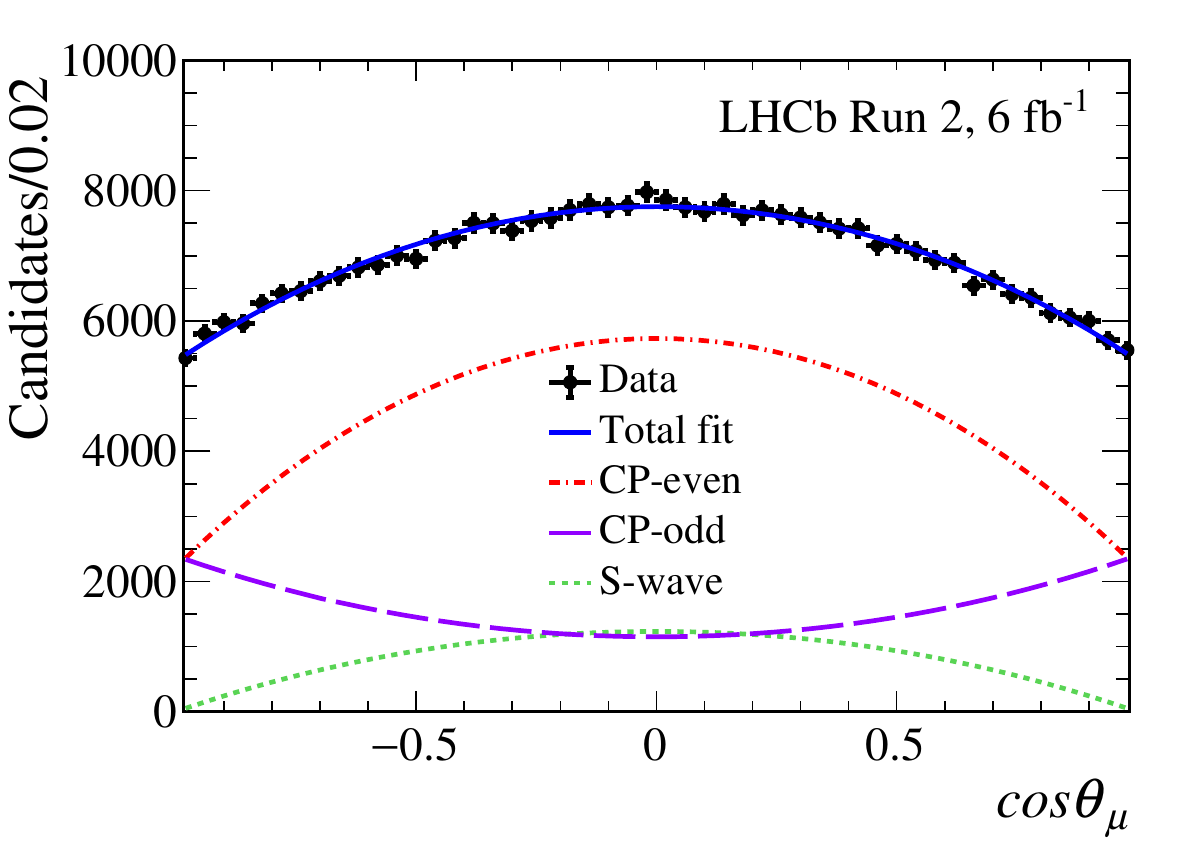}
    \includegraphics[width=0.48\textwidth]{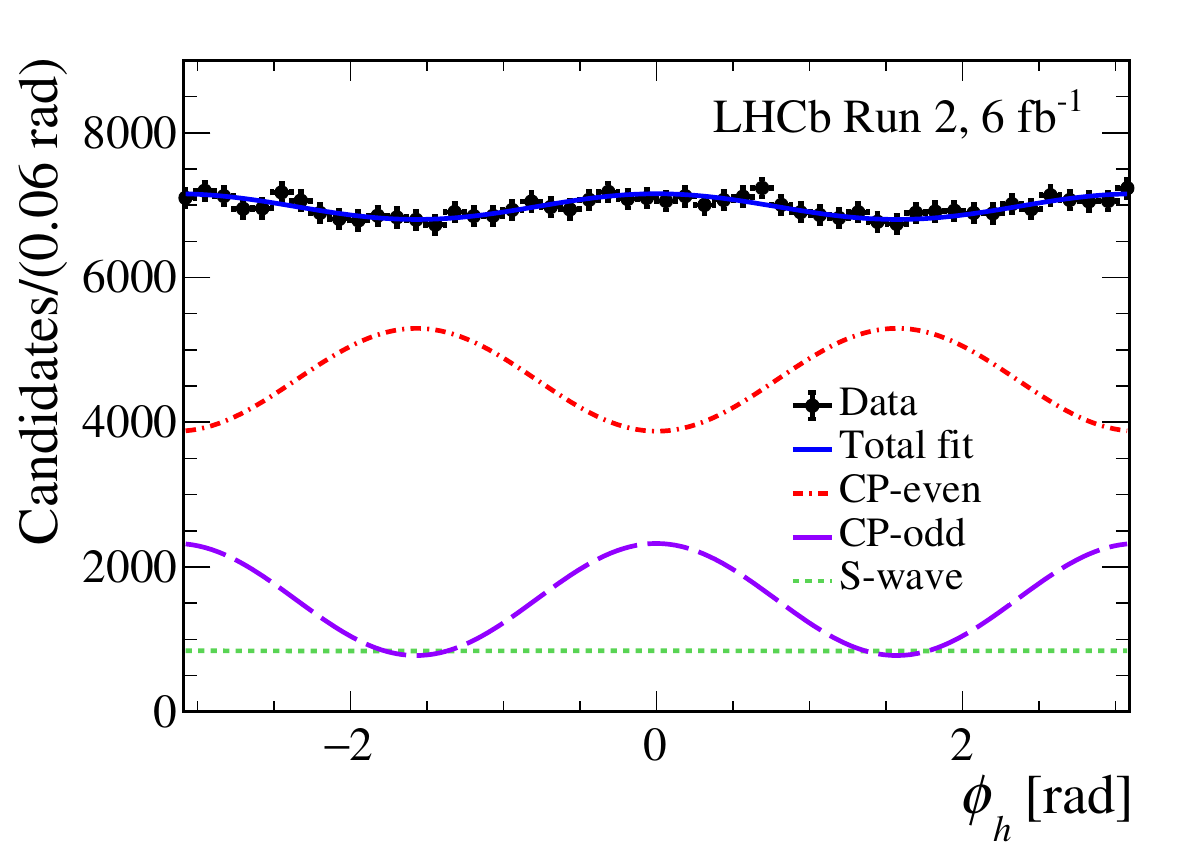}
  \caption{\small Decay-time and decay-angle distributions for background-subtracted \mbox{$B_s^0\to\jpsi\Kp\Km$} decays with the one-dimensional projections of the PDF at the maximum-likelihood point. The data and fit projections for the different samples considered [data-taking year, trigger and tagging categories, $m(\Kp\Km)$ bins] are combined.}
  \label{fig:results_projections}
\end{figure}

The results of the simultaneous maximum likelihood fit to the 48 independent data samples for the nine main physics parameters of interest are given in \tabref{tab:results}. The statistical uncertainties are computed using the profile-likelihood method and cross-checked with the bootstrapping technique~\cite{Efron1979,Christoph2019}.
\begin{table}[htb]
\centering
\caption{Main physics parameters of interest, where the first uncertainty is statistical and the second systematic.~\label{tab:results}}
\begin{tabular}{ll@{\hskip 2pt}l@{\hskip 2pt}ll}
\toprule
Parameter & \multicolumn{3}{c}{Values}\\
\midrule
   $\phi_{s}~[\!\rad]$ & $\phisRunIICV$&$\pm\phisRunIIStatErr$&$\pm\phisRunIISystErr$ \\
     $|\lambda|$           		& $\phantom{+}\lambRunIICV$&$\pm\lambRunIIStatErr$&$\pm\lambRunIISystErr$    \\
     $\Gamma_{s}-\Gamma_{d}~[\!\invps]$ & $\GsGdRunIICV$&$^{\:+\GsGdRunIIStatErrUpper}_{\:-\GsGdRunIIStatErrLower}$&$\pm
     \GsGdRunIISystErr$ \\
     $\Delta\Gamma_{s}~[\!\invps]$ & $\phantom{-}\DGsRunIICV $&$\pm \DGsRunIIStatErr $&$\pm \DGsRunIISystErr$ \\
     $\Delta m_{s}~[\!\invps]$ & $\phantom{+}\DmsRunIICV$&$\pm \DmsRunIIStatErr$&$\pm\DmsRunIISystErr $ \\
     $|A_{\perp}|^2$	& $\phantom{+}\AperpSqRunIICV$&$\pm\AperpSqRunIIStatErr$&$\pm\AperpSqRunIISystErr$  \\
     $|A_0|^2$             		& $\phantom{+}\AzeroSqRunIICV$&$\pm\AzeroSqRunIIStatErr$&$\pm\AzeroSqRunIISystErr$   \\
     $\delta_\perp-\delta_0 ~[\!\rad]$      	& $\phantom{+}\DperpRunIICV$&$^{\:+\DperpRunIIStatErrUpper}_{\:-\DperpRunIIStatErrLower}  $&$\pm \DperpRunIISystErr$ \\
     $\delta_\parallel-\delta_0 ~[\!\rad]$    	& $\phantom{+}\DparRunIICV$&$\pm\DparRunIIStatErr$&$\pm\DparRunIISystErr \ $\\
\bottomrule
\end{tabular}
\end{table}
The background-subtracted data distributions with fit projections are shown in \figref{fig:results_projections}.
The results are in good agreement with the \lhcb Run~1 and 2015--2016 measurements~\cite{LHCb-PAPER-2014-059,LHCb-PAPER-2020-042,LHCb-PAPER-2019-013}. As a cross-check, the analysis is performed on the 2015--2016 data subsample, and the results are consistent with the previous Run~2 measurement~\cite{LHCb-PAPER-2019-013}.
The measurements of $\phi_s$, $\Delta\Gamma_s$, and $\Gamma_s-\Gamma_d$
are the most precise to date and agree with the SM expectations~\cite{UTfit-UT,CKMfitter2015,Artuso:2015swg,Lenz2020}.
No \CP violation in $B_s^0 \to J/\psi \Kp \Km$ decays is found. The value of $\Delta m_{s}$ agrees with the world average~\cite{PDG2022}.
The amplitudes of the {\it S}-wave component are determined in the same fit and summarized in Supplemental Material~\cite{Supplementary}.
Removing the assumption that the \CP-violating parameters $|\lambda|$ and $\phi_s$ are the same for all polarization
states shows no evidence for any polarization dependence, and the corresponding results are summarized in Table~\ref{tab:pol-dep-results}.

\begin{table}[htbp]
    \centering
    \caption{Measured observables in the polarization-dependent fit. The uncertainties are statistical only.}
    \label{tab:pol-dep-results}
    \begin{tabular}{l c} 
    \toprule
       Parameters  &  Values\\ \midrule 
    $\phi_{s}^0~[\!\rad]$   & ~~$-0.034\pm 0.023  $ \\
    $\phi_s^{\parallel}-\phi_{s}^0 ~[\!\rad]$   & ~~$-0.002\pm 0.021 $  \\
    $ \phi_s^{\perp}-\phi_{s}^0~[\!\rad]$  & $-0.001^{\:+0.020}_{\:-0.021} $ \\
    $\phi_s^{S}-\phi_{s}^0~[\!\rad]$ & $\phantom{+}0.022^{\:+0.027}_{\:-0.026} $  \\
    $|\lambda^0|$ & $\phantom{+}0.969 ^{\:+0.025}_{\:-0.024}$   \\
    $|\lambda^{\parallel}/\lambda^0|$ & $\phantom{+}0.982 ^{\:+0.055}_{\:-0.052}$ \\
    $ |\lambda^{\perp}/\lambda^0| $& $\phantom{+}1.107^{\:+0.082}_{\:-0.076} $  \\
    $ |\lambda^{S}/\lambda^0| $& $\phantom{+}1.121 ^{\:+0.084}_{\:-0.078} $  \\ 
    \bottomrule
    \end{tabular}

\end{table}

The total systematic uncertainties shown in \tabref{tab:results} are the quadrature sum of different contributions, described in the following and summarized in Supplemental Material~\cite{Supplementary}. 
The tagging parameters are constrained in the fit, and, therefore, their associated systematic uncertainties contribute to the statistical uncertainty of each physics parameter. This contribution is 0.0025~\rad\ to \phis and 0.0015~$\invps$ to $\Delta m_s$ and is negligible for all other parameters.

The systematic uncertainties related to the mass fit model are estimated by changing the calibration model of the per-candidate mass uncertainty, varying the estimation of misidentified \Lb and mass resolution parameters independently. 
The signal weights are recomputed by varying the fit parameters within their statistical uncertainties.
Since the \sPlot method implicitly relies on factorization
of the discriminating variable, $m(\jpsi K^{+}K^{-})$, and the rest of the observables, a systematic uncertainty is assigned for the small correlation between the $m(\jpsi K^{+}K^{-})$ distribution and the decay time and angles by reevaluating the signal weights in bins of decay time and angles.  
The effect of ignoring the contribution of $B_c^+\to B_s^0 X$ candidates is evaluated with pseudoexperiments by adding about 2\%  ~\cite{LHCb-PAPER-2019-033,LHCb-PAPER-2020-046,Kiselev:2003mp} of $B_c^+\to B_s^0(\to \jpsi\phi)X$ simulated candidates, estimated from the branching fraction and efficiencies. The $B_c^+$ component causes biases in $\Gamma_s$ and $\Delta\Gamma_s$ at the level of about $0.0015~\invps$. These are corrected for in the final fit taking into account the systematic uncertainty on the bias, while minor differences in other parameters are taken as systematic uncertainties. 
The effect of neglecting a possible {\it D}-wave $\Kp\Km$ component is conservatively estimated with pseudoexperiments that contain twice the size of the expected {\it D}-wave contribution from Ref.~\cite{LHCb-PAPER-2012-040}.

The effect due to imperfect removal of ghost tracks~\cite{LHCb-PUB-2021-005} reconstructed with noisy hits is evaluated according to simulation and considered among the systematic uncertainties.
Around 1.4\% of the selected events have multiple candidates; the effect of such cases is considered in systematic uncertainties by choosing one candidate randomly and repeating the fit.
Systematic uncertainties are assigned due to the limited size of the PID calibration samples.
Different models of the {\it S}-wave line shape based on the results in Ref.~\cite{LHCb-PAPER-2013-069} are used to evaluate the $C_{\rm SP}$ factors and assign systematic uncertainties.

A systematic uncertainty for the translation of the decay-time resolution calibration from the control sample to signal is derived using simulation. A minor systematic uncertainty due to non-Gaussian effects in the decay-time resolution is assigned.
Systematic uncertainties accounting for the limited sizes of the calibration sample for the decay-time resolution and the simulated samples for the angular efficiencies are estimated by varying the calibration parameters and efficiencies according to the statistical covariance matrices.  
The effect of ignoring the angular resolutions in the fit is estimated by performing separate fits to the generated and reconstructed angular variables from simulation, and the differences are taken as systematic uncertainties.
The effect of the specific configuration of the gradient-boosted tree method applied to correct the kinematics of simulation in the time and angular efficiency is estimated by applying 100 alternative configurations.

The longitudinal scale of the vertex detector has a relative uncertainty of 0.022\%~\cite{LHCb-DP-2014-001,LHCb-PAPER-2013-006} and a systematic uncertainty is assigned by scaling the track parameters with this uncertainty. 
The systematic uncertainty associated with the track momentum scale calibration is estimated by varying all track momenta by $0.03\%$~\cite{LHCb-2008-037}. 
Possible biases in the fitting procedure and effects of neglecting correlations between the decay-angle and decay-time efficiencies are studied using pseudoexperiments.
The results are found to be stable when repeating the analysis
on subsets of the data, split by the two LHCb magnet polarities, trigger conditions,
year of data taking, number of reconstructed primary vertices, bins of \Bs $p_T$, $\eta$ or tagging categories.

In conclusion, the \CP-violation and decay-width parameters in the decay \mbox{$B_s^0\to\jpsi\Kp\Km$} 
are measured using the full Run~2 dataset collected by the \lhcb experiment. 
 The results are \mbox{$\phi_s = \phisRunIICV\pm\phisRunIIStatErr\pm\phisRunIISystErr$~\rad}, $|\lambda| = \lambRunIICV\pm\lambRunIIStatErr\pm\lambRunIISystErr$, \mbox{$\Gamma_{s}-\Gamma_{d} = \GsGdRunIICV^{\:+\GsGdRunIIStatErrUpper}_{\:-\GsGdRunIIStatErrLower} \pm \GsGdRunIISystErr ~\invps$} and $\Delta \Gamma_{s} = \DGsRunIICV \pm \DGsRunIIStatErr \pm \DGsRunIISystErr~\invps$,
superseding the previous Run~2 \lhcb measurement in the same decay~\cite{LHCb-PAPER-2019-013}.
No evidence for \CP violation is found in the $B_s^0\to\jpsi(\mu^+\mu^{-})\Kp\Km$ decay. The results are consistent with the previous measurements in $B_s^0\to\jpsi(\mu^+\mu^{-})\Kp\Km$~\cite{LHCb-PAPER-2014-059,LHCb-PAPER-2019-013} and $B_s^0\to\jpsi(e^{+}e^{-})\Kp\Km$~\cite{LHCb-PAPER-2020-042} decays and the combination with which yields $\phi_s = \phisAvgJpsiKKCV \pm \phisAvgJpsiKKErr$~\rad\ and $|\lambda| =  \lambAvgJpsiKKCV \pm \lambAvgJpsiKKErr$.  
A combination of all LHCb \phis measurements in \Bs decays via $b\to c\bar{c}s$ transitions~\cite{LHCb-PAPER-2017-008,LHCb-PAPER-2014-051,LHCb-PAPER-2014-019,LHCb-PAPER-2016-027,LHCb-PAPER-2020-042,LHCb-PAPER-2014-059}, 
$B_s^0\to\jpsi(\mu^+\mu^{-})\Kp\Km$ above the $\phi(1020)$ resonance, $B_s^0\to D_s^+ D_s^+$, $B_s^0 \to \jpsi \pip\pim$, $B_s^0 \to \psi(2S)\Kp\Km$ and $B_s^0 \to \jpsi \Kp \Km$, yields $\phi_s = \phisAvgAllCV\pm\phisAvgAllErr$~\rad.
The full fit results and correlations are provided in Supplemental Material~\cite{Supplementary}.
This is the most precise measurement of the \CP-violating phase $\phi_s$ to date and is consistent with SM predictions~\cite{CKMfitter2015,UTfit-UT}.
%The enhanced precision drives the measurement of $\phi_s$ into a new era, necessitating a comprehensive evaluation of the penguin effects.

%\section*{Acknowledgements}
\section*{}
%
% These Acknowledgements valid from 3-May-2019
%
\noindent We express our gratitude to our colleagues in the CERN
accelerator departments for the excellent performance of the LHC. We
thank the technical and administrative staff at the LHCb
institutes.
We acknowledge support from CERN and from the national agencies:
CAPES, CNPq, FAPERJ and FINEP (Brazil); 
MOST and NSFC (China); 
CNRS/IN2P3 (France); 
BMBF, DFG and MPG (Germany); 
INFN (Italy); 
NWO (Netherlands); 
MNiSW and NCN (Poland); 
MCID/IFA (Romania); 
%MSHE (Russia); 
MICINN (Spain); 
SNSF and SER (Switzerland); 
NASU (Ukraine); 
STFC (United Kingdom); 
DOE NP and NSF (USA).
We acknowledge the computing resources that are provided by CERN, IN2P3
(France), KIT and DESY (Germany), INFN (Italy), SURF (Netherlands),
PIC (Spain), GridPP (United Kingdom), 
%RRCKI and Yandex LLC (Russia), 
CSCS (Switzerland), IFIN-HH (Romania), CBPF (Brazil),
Polish WLCG  (Poland) and NERSC (USA).
We are indebted to the communities behind the multiple open-source
software packages on which we depend.
Individual groups or members have received support from
ARC and ARDC (Australia);
Minciencias (Colombia);
AvH Foundation (Germany);
EPLANET, Marie Sk\l{}odowska-Curie Actions, ERC and NextGenerationEU (European Union);
A*MIDEX, ANR, IPhU and Labex P2IO, and R\'{e}gion Auvergne-Rh\^{o}ne-Alpes (France);
Key Research Program of Frontier Sciences of CAS, CAS PIFI, CAS CCEPP, 
Fundamental Research Funds for the Central Universities, 
and Sci. \& Tech. Program of Guangzhou (China);
%Key Research Program of Frontier Sciences of CAS, CAS PIFI,
%Thousand Talents Program, and Sci. \& Tech. Program of Guangzhou (China);
%RFBR, RSF and Yandex LLC (Russia);
GVA, XuntaGal, GENCAT, Inditex, InTalent and Prog.~Atracci\'on Talento, CM (Spain);
SRC (Sweden);
the Leverhulme Trust, the Royal Society
 and UKRI (United Kingdom). \textcolor{white}{Victoria Bumma}

\newpage

%TC:ignore

\bibliographystyle{LHCb}
\bibliography{main,standard,LHCb-PAPER,LHCb-CONF,LHCb-DP,LHCb-TDR}

\clearpage
\appendix
% This should be taken out in the final paper
%\input{supplementary}
% LHCb collaboration author list
% Data extracted on July 18th, 2023 at 5:01pm for paper reference LHCb-PAPER-2023-016
\newpage
\centerline
{\large\bf LHCb Collaboration}
\begin
{flushleft}
\small
R.~Aaij$^{33}$\lhcborcid{0000-0003-0533-1952},
A.S.W.~Abdelmotteleb$^{52}$\lhcborcid{0000-0001-7905-0542},
C.~Abellan~Beteta$^{46}$,
F.~Abudin{\'e}n$^{52}$\lhcborcid{0000-0002-6737-3528},
T.~Ackernley$^{56}$\lhcborcid{0000-0002-5951-3498},
B.~Adeva$^{42}$\lhcborcid{0000-0001-9756-3712},
M.~Adinolfi$^{50}$\lhcborcid{0000-0002-1326-1264},
P.~Adlarson$^{78}$\lhcborcid{0000-0001-6280-3851},
H.~Afsharnia$^{10}$,
C.~Agapopoulou$^{44}$\lhcborcid{0000-0002-2368-0147}
% ...

{\footnotesize \it

% ...
$^{10}$Universit{\'e} Clermont Auvergne, CNRS/IN2P3, LPC, Clermont-Ferrand, France\\
% ...
$^{33}$Nikhef National Institute for Subatomic Physics, Amsterdam, Netherlands\\
% ...
$^{42}$Instituto Galego de F{\'\i}sica de Altas Enerx{\'\i}as (IGFAE), Universidade de Santiago de Compostela, Santiago de Compostela, Spain\\
% ...
$^{44}$European Organization for Nuclear Research (CERN), Geneva, Switzerland\\
% ...
$^{46}$Physik-Institut, Universit{\"a}t Z{\"u}rich, Z{\"u}rich, Switzerland\\
% ...
$^{50}$H.H. Wills Physics Laboratory, University of Bristol, Bristol, United Kingdom\\
% ...
$^{52}$Department of Physics, University of Warwick, Coventry, United Kingdom\\
% ...
$^{55}$School of Physics and Astronomy, University of Glasgow, Glasgow, United Kingdom\\
$^{56}$Oliver Lodge Laboratory, University of Liverpool, Liverpool, United Kingdom\\
% ...
$^{78}$Department of Physics and Astronomy, Uppsala University, Uppsala, Sweden, associated to $^{55}$\\
% ...
}
\end{flushleft}
\newpage
\clearpage
%\input{JustificationPRL}
%TC:endignore

\newcommand{\detailtexcount}[1]{%
  \immediate\write18{texcount -merge -sum -q #1.tex output.bbl > #1.wcdetail }%
  \verbatiminput{#1.wcdetail}%
}

\newcommand{\quickwordcount}[1]{%
  \immediate\write18{texcount -1 -sum -merge -q #1.tex output.bbl > #1-words.sum }%
  \input{#1-words.sum} words%
}

\newcommand{\quickcharcount}[1]{%
  \immediate\write18{texcount -1 -sum -merge -char -q #1.tex output.bbl > #1-chars.sum }%
  \input{#1-chars.sum} characters (not including spaces)%
}

%TC:ignore
%\detailtexcount{main}
%\quickwordcount{main}
%\quickcharcount{main}
%TC:endignore

\end{document}